\documentclass[twocolumn,prX]{revtex4}
\begin{document}
\draft
\title{
Fluctuation-dissipation theorem for thermo-refractive noise.
}
\author{  Yuri Levin
}
\address{Leiden University, Leiden Observatory and Lorentz Institute, Niels Bohrweg 2,
2300 RA Leiden, the Netherlands}
\date{\today}
\begin{abstract}
We introduce a simple prescription for calculating the spectra
of thermal fluctuations of temperature-dependent quantities
of the form $\hat{\delta T}(t)=\int d^3\vec{r} \delta T(\vec{r},t) q(\vec{r})$. Here
$T(\vec{r}, t)$ is the local temperature at location $\vec{r}$ and time $t$, 
and $q(\vec{r})$ is an arbitrary function. As an example of a possible
application, we compute the spectrum of thermo-refractive coating noise in LIGO,
and find a complete agreement with the previous calculation of Braginsky,
Gorodetsky and Vyatchanin. Our method has computational advantage, especially
for non-regular or non-symmetric geometries, and for the cases where 
$q(\vec{r})$ is non-negligible in a significant fraction of the total volume.  
\end{abstract}
\maketitle

\section{introduction and main results}
The theory of time-dependent thermodynamical fluctuations has been extensively 
developed for the past
century. One of the fundamental results in this field is the Fluctuation-dissipation Theorem
(FdT), which was originally formulated by Callen and Welton in 1951 \cite{callenwelton}. Several different formulations
of the FdT have been introduced since then, see \cite{beenaker} for a review. In this short paper
we develop a formulation which is suitable for calculating the spectra $S_{\hat{\delta T}}(\omega)$ 
of thermal
fluctuations of temperature-dependent quantities
of the form 
\begin{equation}
\hat{\delta T}=\int d^3\vec{r} \delta T(\vec{r},t) q(\vec{r}).
\label{hatT}
\end{equation}
Here
$\delta T(\vec{r},t)$ is the local temperature at location $\vec{r}$ and time $t$, 
and $q(\vec{r})$ is an arbitrary function. 

Computations of this kind may be  relevant for the design of 
interferometric gravitational-wave detectors 
like the Laser Interferometer Gravitational Wave Observatory
(LIGO).
Random thermal fluctuations are expected to be the dominant noise source
for LIGO at  frequencies between $10$ and $100$ Hz; see, e.g., \cite{ligovirgo}. 
This  noise is mostly 
due to the thermal motion of the LIGO mirror surfaces, and has been 
extensively studied both theoretically and experimentally by several groups around the
world; see, e.g.,  \cite{black} and references therein. 
More recently Braginsky, Gorodetsky, and Vyatchanin (\cite{BGV}, hereafter BGV)
have identified a different kind of thermal noise, which they called the thermo-refractive (TR) noise.
The TR noise can be understood as follows: 

Consider a laser beam which passes through  (the part of) one of
LIGO test masses. 
The index of refraction of the test-mass material
is strongly temperature-dependent. Thus local 
thermodynamical fluctuation in the temperature $\delta T(\vec{r})$ result in the
fluctuation of the overall phase of the laser beam, as measured by an
interferometric experiment. For small expected temperature fluctuations,
the variation of the phase is of the same form as in Eq.~(\ref{hatT}). 
Thus a computation of the thermo-refractive noise amounts to that of the
fluctuations in $\hat{\delta T}$.
BGV have computed the TR noise for a  plane-parallel geometry of the
thin mirror coating. Their computation was based on the multi-dimensional Langevin equations
and involved a calculation of the spacial temperature correlation functions.
While BGV approach works fine for simple geometries, it becomes numerically
cumbersome in more general geometries, i.e.~in a case when the light goes
through one of the test masses and the light-beam radius is comparable to that 
of the test mass.

By contrast, our computational approach will be based on a direct application
of the FdT, in the spirit of our earlier treatment of the
mechanical thermal noise (\cite{L98}, hereafter L98). 
We will show that in order to compute the fluctuations
in $\hat{T}$, one needs to perform the following mental experiment consisting of 3 steps:

1. Periodically inject entropy into the medium, with the volume density of the
entropy injection given by
\begin{equation}
{\delta s(\vec{r})\over dV}=F_0 \cos(\omega t)q(\vec{r}),
\label{ds}
\end{equation}
where $F_0$ is an arbitrarily  small constant.

2. Track all thermal relaxation processes in the system (e.g., the 
heat exchange between different parts of the system) which
occur as a result of the periodic entropy injection. Calculate the total entropy
production rate and hence the total dissipated power $W_{\rm diss}$ which occurs as
a result of the thermal relaxation.

3. Evaluate the spectral density of fluctuations in $\hat{\delta T}$ 
via the formula below:
\begin{equation}
S_{\hat{\delta T}}(f)={8k_BT\over \omega^2}{W_{\rm diss}\over F_0^2};
\label{shat}
\end{equation}
cf.~Eq.~(1) of L98. Here $\omega=2\pi f$.

In the following section we give a
 proof of the above prescription. In section 3,
as an illustration of the method,
we compute the thermo-refractive coating noise in
the BGV geometry. We conclude with some brief general discussion in section 4.

\section{Proof}
It may well be possible to verify Eq.~(\ref{shat}) by appealing to one of the existing 
formulations of the FdT, which treat the fluctuations of  generalized thermodynamic variable
(see \cite{beenaker}). However, the gravitational-wave community 
(including this author) is much more familiar
with the FdT for a generalized {\it mechanical} coordinate of the system, as given by
Callen and Welton in 1951 \cite{callenwelton}. We thus think it is instructive to construct a proof of
Eq.~(\ref{shat}) using a mechanical coordinate.

For this, we 
mentally introduce a set of non-intrusive mechanical thermometers into the system.
Our thermometers are localized ensembles
of identical harmonic oscillators, which are assumed to\newline 
(a) be sufficiently densely packed into the system (we will make this more precise shortly),\newline
(b) have total mass and heat capacity which are vanishingly small compared with those of the
original system (for the latter it is sufficient to assume
that the number of thermometers is much smaller than the number of particles
in the system), \newline 
(c) have a proper angular frequency $\omega_0$ which
is much higher than the angular frequencies $2\pi f$ 
at which $S_{\hat{\delta T}}(f)$ is computed, and \newline
(d) be thermally coupled to the system on a timescale much shorter
than $1/f$.

Consider now an operator
\begin{equation}
\hat{T}_1=\Sigma_i x_i^2,
\label{T1}
\end{equation}
where $x_i$ is the displacement of the $i$'th oscillator. For densely packed
oscillators, the equation above can be written as
\begin{equation}
\hat{T}_1=\int d^3r n(\vec{r}) <x^2(\vec{r})>,
\label{T11}
\end{equation}
where $n(\vec{r})$ is the spacial density of the oscillators, 
and $<x^2(\vec{r})>$ is the average $x^2$ at a radius $\vec{r}$.
Note that as the number of oscillators increases, the Eq.~(\ref{T11})
becomes better defined and more precise. For sufficiently dense
packing,
\begin{equation}
<x^2(\vec{r})>={k_B\over m\omega_0^2}T(\vec{r}),
\label{x2}
\end{equation}
where $k_B$ is the Boltzmann constant, and $m$ is the oscillator mass.
Here, when we say ``sufficiently dense'', we mean that in a minimum
volume for which the notion of local temperature $T(\vec{r})$ 
is meaningful, there should be a number of oscillators $\gg 1$. If
such volume has $N_1\gg 1$ particles, then it should have $N_2$ oscillators where
$N_1\gg N_2\gg 1$. 

We now choose the density of the oscillators be
\begin{equation} 
n(\vec{r})={m\omega_0^2\over k_B}q(\vec{r}),
\label{n}
\end{equation}
where $q(\vec{r})$ is the form factor from Eq.~(\ref{hatT}); we can always
rescale $q(\vec{r})$ so that the oscillators are sufficiently densely
packed. With this choice of $n(\vec{r})$, the dynamical variable
$\hat{T}_1$ closely tracks the thermodynamical variable $\hat{T}$. 

We can now use the Callen-Welton formulation of the FdT to find the fluctuation
$S_{\hat{T}}(f)$
in $\hat{T}_1$. We follow closely the steps described in L98:\newline

{\bf Step 1.} We introduce a periodic perturbation of the form of the interaction Hamiltonian
\begin{equation}
H_{\rm int}=-F_0 \cos(2\pi ft) \hat{T}_1,
\label{hint}
\end{equation}
and consider response of the system to this perturbation. From Eq.~(\ref{T1})
we see that physically such perturbation amounts to a periodic change in the
rigidity of each of the oscillator, or, equivalently, in a periodic change $\delta \omega_0$
in the oscillator proper frequency:
\begin{equation}
{\delta\omega_0\over \omega_0}={F_0\cos(2\pi ft)\over m\omega_0^2}.
\label{deltaomega0}
\end{equation}
If the oscillators were not thermally coupled to the system, their energy 
would track adiabatically the change in the proper frequency, 
$\delta E/E=\delta\omega_0/\omega_0$. Once the thermal coupling is included, their
energy change is given by
\begin{equation}
{\delta E_i\over E_i}={\delta\omega_0\over \omega_0}-{\delta Q_i\over E_i},
\label{delEi}
\end{equation}
where $\delta Q_i$ is the energy input  from the $i$'th oscillator to the local 
thermal bath.
By construction the oscillators' thermal coupling
to the system is much more rapid than $1/f$, and thus on average their energy does not
change (it remains $k_BT$). Therefore, once averaged over the local volume $dV$,
Eq.~(\ref{delEi}) gives
\begin{equation}
\delta Q={F_0\cos(2\pi ft)\over m\omega_0^2}\times k_B T n(\vec{r})\times dV.
\label{deltaQ1}
\end{equation} 
We now substitute Eq.~(\ref{n}) into the above equation, and get
\begin{equation}
{\delta s\over dV}={1\over T}{\delta Q\over dV}=F_0\cos(2\pi ft) q(\vec{r}).
\label{deltaQ2}
\end{equation}
This is the density of the local entropy injection as a result of the periodic
perturbation driven by $H_{\rm int}$.\newline

{\bf Step 2.} We compute the total power $W_{\rm diss}$ which is dissipated in the system
as a result of the periodic forcing. This means that we compute the total entropy production
in the system; from Eq.~(\ref{deltaQ2}) we see that it is zero to first order in $F_0$.
Not so to the second order: the entropy injection leads to temperature inhomogeneities
which in turn lead to thermal relaxation processes, i.e. to the heat exchange between
different parts of the system. The thermal relaxation leads to the net entropy production
rate which is second order in $F_0$. \newline

{\bf Step 3.} We compute 
$S_{\hat{\delta T}}(f)$ by using Eq.~(\ref{shat}). This completes justification for
and physical description of the computational procedure outlined in the introduction.
In next section we work out a practical example, as a useful crosscheck of our approach.

\section{Thermo-refractive noise in LIGO mirror coating}
BGV studied the thermo-refractive noise in the thin optical coating of LIGO mirrors.
We re-derive their result using the direct approach developed above.

As BGV have explained, the relevant variable for the coating thermo-refractive fluctuation
is
\begin{equation}
\hat{\delta T}={1\over \pi r_0^2 l}\int_{-\infty}^{\infty}dx dy\int_0^\infty dz \delta T(\vec{r},t)
e^{-(x^2+y^2)/r_0^2}e^{-z/l},
\label{hatt1}
\end{equation}
where $r_0$ is the effective beam size, $l$ is the effective coating thickness, $x$ and $y$ are the
rectangular coordinates along the mirror face, and $z$ is the coordinate along the beam,
chosen so that $z=0$ at the coating outside boundary. Within our approach, we need
to, as a thought experiment,  inject an oscillating entropy perturbation of the form in Eq.~(\ref{ds}), where
\begin{equation}
q(\vec{r})={1\over \pi r_0^2 l}e^{-(x^2+y^2)/r_0^2}e^{-z/l}.
\end{equation}
The heat injection leads to periodic temperature perturbation $\delta T(\vec{r})$, which
we compute below. We can then find the dissipated power [see Eq.~(35.1) of Landau and Lifshitz \cite{landau}
and Eq.~(5) of \cite{liu}]:
\begin{equation}
W_{\rm diss}=\int d^3r {\kappa\over T}\left<\left(\nabla \delta T\right)^2\right>,
\label{wdiss2}
\end{equation}
where $\kappa$ is the thermal 
conductivity, and $<...>$ stands for averaging over the oscillation
period $2\pi/\omega$. 

As BGV have noted, the thermal diffusion lengthscale $l_{\rm th}=\sqrt{\kappa/(C\rho\omega)}$ 
satisfies the following inequality:
\begin{equation}
r_0\gg\l_{\rm th}\gg l.
\end{equation}
Therefore, (1) the heat diffusion is almost exclusively in the $z$ direction, and
(2) we can consider all of the entropy to be injected at the outer
coating boundary, with the surface density
\begin{equation}
{\delta s\over dA}={F_0\cos(\omega t)\over \pi r_0^2} e^{-(x^2+y^2)/r_0^2}.
\label{dsda}
\end{equation}
The temperature perturbation satisfies the diffusion equation
\begin{equation}
{\partial \delta T\over \partial t}=D{\partial^2 \delta T\over \partial z^2},
\label{diffusion}
\end{equation}
and the boundary condition
\begin{equation}
-\kappa \left({\partial \delta T\over \partial z}\right)_{z=0}=T {\partial\over\partial t}
{\delta s\over dA}.
\label{boudary}
\end{equation}
The latter simply states that all the heat injected at the boundary is transported
inwards\footnote{Some of this heat is radiated from the surface, and in principle 
this effect should be taken into account. However, one can easily show that diffusion is
by far more efficient in transporting heat than radiation.}.
 Here $D=\kappa/(\rho C)$,
where $\rho$ is the density and $C$ is the heat capacity.

It is now straightforward to find $\delta T$ and its gradient:
\begin{equation}
{\partial T\over \partial z}=
{TF_0\omega\over \kappa\pi r_0^2}\exp\left(-\sqrt{\omega\over 2D}z\right)
  \sin \left(\sqrt{\omega\over 2D}z-\omega t\right)e^{-(x^2+y^2)/r_0^2}.
\label{dtdz}
\end{equation}
We can now find the dissipated power:
\begin{equation}
W_{\rm diss}\simeq\int d^3r {\kappa \over T}
 \left<\left({\partial \delta T\over \partial z}\right)^2
\right>={F_0^2\omega^2T\over 4\pi r_0^2 \kappa}\sqrt{D\over 2\omega}.
\label{wdiss5}
\end{equation}
Finally, from Eq.~(\ref{shat}), we get
\begin{equation}
S_{\hat{\delta T}}(f)={\sqrt{2}k_B T^2\over \pi r_0^2 \sqrt{\omega C \rho\kappa}}.
\label{Sfinal}
\end{equation}
This result is identical to Eq.~(9) of BGV.
 
%coating thickness. Thus in our mental experiment the coating is isothermal. Furthermore,
%since $l_{\rm th}$ is much smaller than the beam size, the heat diffusion in the direction
%transverse to the beam is much smaller than that in the direction along the beam.
%Thus the diffusion is essentially one-dimentional, and the diffusion equation can be written as
%\begin{equation}

%\subsection{Isothermal small body}

\section{Discussion}
Our direct approach to the thermo-refractive noise has 2 advantages, as compared to
the  previously used methods:\newline

1. It is computationally easier, especially when the form-factor $q(\vec{r})$ is significant over a
large fraction of the system's volume. For example, a broad beam of light passing through a LIGO
test mass will be sensitive to the  thermo-refractive fluctuations in a large
fraction of the test mass' volume.\newline

2. In elucidates geometries and material properties which may make the thermo-refractive noise high.
The thermal relaxation which is induced in the process of our mental experiment
depends on the spacial variation of the form-factor $q(\vec{r})$ and (less obviously) on that
of the heat capacity.\newline
Future work may include unified simultaneous modelling of thermo-refractive and thermoelastic noises,
since they both originate from $T$-fluctuations and may be highly correlated. 

\section{acknowledgements}
We thank Carlo Beenakker for useful discussions.

%\newpage
%\bibitem{callenwelton} H.\ B.\ Callen and T.\ A.\ Welton, {\it Phys. Rev.} {\bf 83}, 34-50 (1951).
%\begin{figure}
%\caption[]{
%Identical defects A and B create fluctuating strees in different parts
%of the test mass. The stress created by defect A will influence 
%the phase shift of the laser beam readout more than the stress created by 
%defect B, although both A and B make identical contributions to 
%$Q$'s  of the test mass's elastic modes.
%}
%\label{fig:readout}
%\end{figure}

\begin{references}


\bibitem{callenwelton} H.\ B.\ Callen and T.\ A.\ Welton, {Phys. Rev.} {\bf 83}, 34-50 (1951).

\bibitem{beenaker}  S.\ R.\ de Groot, and P.\ Mazur, {\it Non-equilibrium 
Thermodynamics} (Dover, 1984)

\bibitem{ligovirgo}  A.\ Abramovici {\it et. al.}, Science, {\bf 256}, 325 (1992);
C.\ Baradaschia {\it et. al.}, Nucl. Instrum \& Methods, {\bf A289}, 518 (1990).

\bibitem{black} G.\ M.\ Harry, {\it et al.}, Class.~\& Quantum Gravity, 24, 405
(2007), and references therein. 
%\bibitem{saulson} P.\ R.\ Saulson, {\it Phys. Rev. D}, {\bf 42}, 2437-2445 (1990).

%\bibitem{raab}  A.\ Gillespie and F.\ Raab, {\it Phys. Rev. D}, {\bf 52}, 577-585 (1995).

%\bibitem{vinet} F.\ Bondu and J.\ Y.\ Vinet, {\it Phys. Lett. A}, {\bf 198 (2)}, 74-78 (1995).

%\bibitem{gonzales} G.\ I.\ Gonzalez and  P.\ R.\ Saulson, {\it J. Accoust. Soc. Am}, {\bf 96},
%                     207-212 (1994). 

%\bibitem{raab1} A.\ Gillespie and F.\ Raab, {\it Phys. Lett. A}, {\bf 178}, 357-363 (1993).

\bibitem{BGV} V.\ B.\ Braginsky, M.\ Gorodetsky, and S.\ P.\ Vyatchanin, 
      {Phys. Lett. A}, 271, 303 (2000), BGV in the text.
%                    references therein; private communications. 

\bibitem{L98} Y.\ Levin, { Phys.~Rev.~D}, 57, 659 (1998), L98 in the text.

\bibitem{landau} L.\ D.\ Landau, and E.\ M.\ Lifshitz, {\it 
Theory of Elasticity},
third edition (Pergamon, Oxford, 1986)

\bibitem{liu} Y.\ Liu, and K.\ S.\ Thorne, Phys.~Rev.~D, 6212002 (2000)

%\bibitem{landau} Equation (8.19) of L.\ D.\ Landau and E.\ M.\ Lifshitz, {\it Theory of Elasticity}
%                  (Pergamon Press, New York, 1986).


%\bibitem{callenwelton} H.\ B.\ Callen and T.\ A.\ Welton, {\it Phys. Rev.} {\bf 83}, 34-50 (1951).

\end{references}
\end{document}